\documentclass[manuscript,screen]{acmart}
\usepackage{tabu}
\usepackage{multirow}
\usepackage{subcaption}
\usepackage{caption}
\usepackage{graphicx}
\usepackage{footnote}  
\usepackage{float}
\AtBeginDocument{%
  \providecommand\BibTeX{{%
    \normalfont B\kern-0.5em{\scshape i\kern-0.25em b}\kern-0.8em\TeX}}}

\setcopyright{acmcopyright}
\copyrightyear{2023}
\acmYear{2023}
\setcopyright{acmcopyright}\acmConference[ACM Trans. Spatial Algorithms Syst.]{ACM Transactions on Spatial Algorithms and Systems}{April 1, 2023}{Seattle, WA, USA}
\acmBooktitle{ACM Transactions on Spatial Algorithms and Systems}
\acmPrice{15.00}
\acmDOI{XXXXXXX.XXXXXXX}
\acmISBN{978-1-4503-XXXX-X/18/06}

\begin{document}

\title{Spatial Computing Opportunities in Biomedical Decision Support: The Atlas-EHR Vision}

\author{Majid Farhadloo}
\email{farha043@umn.edu}
\affiliation{%
  \institution{Dept. of Computer Science and Engineering, University of Minnesota, Twin Cities}
  \country{USA}
}

\author{Arun Sharma}
\email{sharm485@umn.edu@umn.edu}
\affiliation{%
  \institution{Dept. of Computer Science and Engineering, University of Minnesota, Twin Cities}
  \country{USA}
}
\author{Shashi Shekhar}
\email{shekhar@umn.edu}
\affiliation{%
  \institution{Dept. of Computer Science and Engineering, University of Minnesota, Twin Cities}
  \country{USA}
}

\author{Svetomir N. Markovic}
\email{markovic.svetomir@mayo.edu}
\affiliation{%
  \institution{Dept. of Oncology, Mayo Clinic}
  \country{USA}
}


\renewcommand{\shortauthors}{Farhadloo, et al.}
\renewcommand{\shorttitle}{Spatial Opportunities in Biomedical Decision Support: The Atlas-EHR Vision}

\begin{abstract}
We consider the problem of reducing the time needed by healthcare professionals to understand patient medical history via the next generation of biomedical decision support. This problem is societally important because it has the potential to improve healthcare quality and patient outcomes. However, navigating electronic health records is challenging due to the high patient-doctor ratios, potentially long medical histories, the urgency of treatment for some medical conditions, and patient variability. The current electronic health record systems provides only a longitudinal view of patient medical history, which is time-consuming to browse, and doctors often need to engage nurses, residents, and others for initial analysis. To overcome this limitation, we envision an alternative spatial representation of patients’ histories (e.g., electronic health records (EHRs)) and other biomedical data in the form of Atlas-EHR. Just like Google Maps allows a global, national, regional, and local view, the Atlas-EHR may start with an overview of the patient’s anatomy and history before drilling down to spatially anatomical sub-systems, their individual components, or sub-components. Atlas-EHR presents a compelling opportunity for spatial computing since healthcare is almost a fifth of the US economy. However, the traditional spatial computing designed for geographic use cases (e.g., navigation, land-surveys, mapping) faces many hurdles in the biomedical domain. This paper presents a number of open research questions under this theme in five broad areas of spatial computing. 
\end{abstract}

\maketitle
\section{Introduction}
Electronic health records (EHRs) and other biomedical data are critical sources of information for healthcare professionals, providing them with valuable insights into patient health and medical history. However, the sheer amount of data can be time-consuming for healthcare professionals to navigate. Toward this goal, we envision the next generation of biomedical decision support to reduce the time spent by healthcare professionals examining and understanding medical history from EHRs data.  

\textbf{Societal Importance:} This problem is societally important since it helps healthcare professionals make efficient and accurate informed decisions. This may ultimately improve the quality of care and patient outcomes. Also, by rethinking how medical professionals access and interpret patient medical histories, they may have more time to spend and interact with patients, resulting in improved healthcare standards. 

\textbf{Challenges:} However, the problem is challenging due to several reasons. First, patient visits are short (e.g., around 15 minutes). In addition, a patient’s medical history can be long spanning decades. Furthermore, many primary care doctors are responsible for hundreds and thousands of patients. Additionally, changes in medical guidelines and practices introduce additional complexities. Some medical conditions also require treatment urgently. Further, some medical decisions are sensitive to patient demographics and underlying medical conditions. Moreover, ensuring explainability, fairness, accountability, security, and trustworthiness in EHR systems is paramount challenge for the advancement of biomedical decision support. Explainability improves patient care by enabling healthcare providers to understand and trust the system’s recommendations. Fairness is vital to mitigate biases that could influence treatment outcomes, especially considering the diverse interplay of spatial and location variability, demographics, genetics, and lifestyle. Accountability and security are fundamental for safeguarding sensitive patient information to maintain privacy and adhering to healthcare regulations.  Trustworthy EHR systems that embody these attributes are indispensable for delivering informed and equitable biomedical decision support.

\textbf{Related Work:} Popular EHR systems (e.g., Epic Systems \cite{epic-systems}) provide a longitudinal view of a patient's history, which makes it time-consuming for medical professionals to navigate through (See Figure \ref{old_ehr} for illustration). This representation does not allow medical professionals to explore the interdependencies of EHRs data generated through different sensing modalities over various timescales spanning decades from the whole body to their components or sub-components. Early spatial vision approach includes  inner space \cite{oliver2011geography}. The paper discussed harnessing traditional medical imaging (e.g., X-ray, MRI) data and the concept of spatial networks to build a spatiotemporal database. Their spatial framework was supposed to answer long-term questions about how a disease would progress or the comparative effectiveness of a therapeutic intervention. Challenges they identified at the time included the lack of a spatial reference frame for the human body, how to determine the location in the body, routing in a continuous space, observing change across snapshots, and scalability. More recent work \cite{wolfson2018understanding} explored the human brain as a spatiotemporal object. It highlighted the challenges of GIS-inspired methods (e.g., spatial aggregation for brain parcellations) to study the brain structure and signals communicated with it, which dynamically change depending on the mental task or the function performed by the person. Other literature work on the human cell atlas \cite{humanCellAtlas2024, regev2017human, rozenblatt2017human, haniffa2021roadmap, travaglini2020molecular, han2020construction, cao2020human, schiller2019human, yuan2016spatial} and the human biomolecular atlas program \cite{hubmap2019human, jain2023advances, vento2023cell, lake2021atlas, greenbaum2023spatially, ghose20233d} along with the early emergence of digitized human anatomies \cite{nationalgeographic2021, lozano20173d, digitizedHumanaAnatomy2023} to form a reference map as a basis for understanding human health to improve diagnosis, prognosis, monitoring, and treatment of disease. However, these studies are either limited to specific types of EHR data lack a comprehensive perspective regarding the electronic health record data across the human body as a system of systems \cite{ackoff1971towards, jackson1984towards} through the lens of spatial computing research. A more detailed description of the human cell atlas initiative is provided in Section \ref{emergin_spatial_data}.

\begin{figure}
    \centering
    \includegraphics[width=\linewidth, height=10cm]{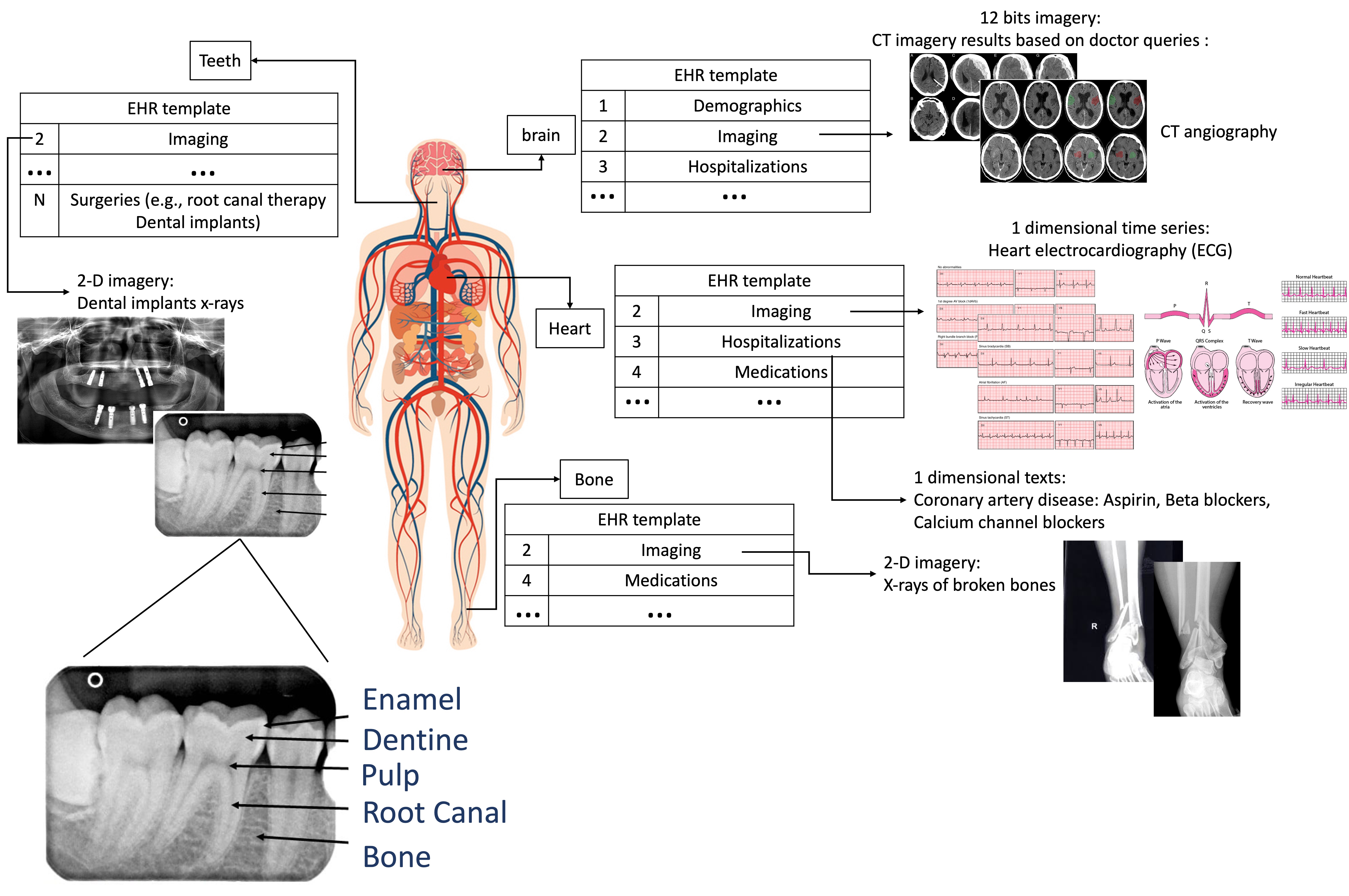}
    \caption{A spatially-organized map of electronic health records}
     \label{atlas-ehr}
\end{figure}

\textbf{Our Atlas-EHR Vision:} 
We envision an alternative, unified spatial framework for electronic health records (EHRs) and other biomedical data, which we call Atlas-EHR. This framework would empower healthcare professionals and potentially other end-users to explore the human body by overlaying medical records onto an interactive 4-dimensional anatomy-based display, encompassing space and time, and covering multiple levels of detail from organ-level associations to sub-cellular interactions. These spatial representations, akin to digital maps like Google Maps, allows users to navigate from a global view of the patient’s anatomy and history to more detailed views of anatomical sub-systems and their individual components or sub-components. Figure \ref{atlas-ehr} illustrates a perspective of Atlas-EHR, where a patient’s medical records are spatially organized based on their anatomy, integrating various structures from organs to tissues and cells. For example, it shows how medical imaging (e.g., heart electrocardiogram time series data) and clinical notes (e.g., medical prescriptions) can be spatially grouped based on relevant organs, such as the heart.

Furthermore, Atlas-EHR extends beyond cartography and geo-visualization, encompassing areas such as spatial data science \cite{shekhar2015spatial, atluri2018spatio, shekhar2020spatial, shekhar2011identifying} and spatial database management systems \cite{eldawy2015spatialhadoop, eldawy2015era} to leverage recent advancements in computer infrastructures (e.g., graphical processing units (GPUs) \cite{mittal2015survey, pandey2022transformational}), cloud computing \cite{marinescu2022cloud, gonzalez2015cloud}, edge computing \cite{cao2020overview, chen2019deep, shi2016edge}, and developments in artificial intelligence and machine learning \cite{ongsulee2017artificial, lecun2015deep, helm2020machine, raschka2020machine, rubinger2023machine}. Moreover, Atlas-EHR leverages immersive technologies like augmented reality (AR) and virtual reality (VR). The applications of immersive technologies may be varied, ranging from educational uses to surgical simulations \cite{WylieWong2023, DonnaMarbury2022, DonnaMarbury2020, NathanEddy2020}. An illustrative example is a real-time AR captioning tool with object detection capabilities, which can elucidate the steps of a medical procedure and emphasize different anatomical components, like substructures within the human body. For instance, consider an x-ray of teeth shown on the left side of Figure \ref{atlas-ehr}: an AR system could highlight various parts of the dental structure and, based on the severity of tooth decay, might suggest interventions such as fillings or root canal treatments.


We envision healthcare providers being able to perceive not only the 'what' but also the 'where' and 'how' of patient data—elements often overlooked in conventional methods of data presentation. Spatial computing has the potential to transform EHRs into dynamic, interactive maps of human health. Atlas-EHR represents not only the current state of a patient's health but also provides predictive insights based on observed spatial and temporal patterns. This endeavor has the potential to bridge the gap between the spatial computing research community and the biomedical domain, enabling medical professionals to easily navigate EHRs and other biomedical data using a spatial approach. 
In the following sections, we highlight some of the recent breakthroughs in biomedical decision support that were not possible a decade ago and explain how these advancements can be integrated into our Atlas-EHR vision through the lens of spatial computing to improve healthcare quality. Atlas-EHR poses many challenges and research opportunities for current spatial computing, which is traditionally designed for geographic use cases (e.g., navigation, land-survey, mapping).

\textbf{This paper makes the following contributions:}

\begin{itemize}
    \item It envisions a spatially-organized map of electronic health record data for reducing the time needed by healthcare to understand patient history via the next generation of biomedical decision support. Our vision is more comprehensive (e.g., EHR) than related works (e.g., inner space \cite{oliver2011geography}). 
    \item It challenges and opportunities for five broad areas of spatial computing research to explore the human body. 
\end{itemize}

\textbf{Scope:} This is a vision paper; hence prototyping, implementation, and experimental studies fall outside the scope of this article. A detailed description of spatial computing research in geographic space is also outside the scope of this paper. Interested audiences can refer to available books and surveys \cite{shekhar2020spatial, shekhar2011identifying, golmohammadi2018introduction, xie2017transdisciplinary}. This paper also does not address biological interpretation or mechanisms of domain-related applications in clinical trials. We use the term "inner space" \footnote{Innerspace \cite{innerspaceMovie} is a sci-fi movie directed by Spielberg that retells the story of Fantastic Voyage \cite{fantasticVoyage}, in which a miniature submarine traverses the human body to examine the brain, blood arteries, eyes, and ears.}" term to refer to the human body and its inner parts as a fine biomedical scale for distinguishing from the geographic data used in traditional spatial computing (e.g., GPS, location-based services). Table \ref{tab:taxanomy} provides a taxonomy of the spatial computing methods discussed in this paper.

\textbf{Outline: } The rest of this paper is organized as follows: Section \ref{c&o} provides an overview of recent innovations and outlines several challenges and opportunities in the field of spatial computing. Section \ref{emergin_spatial_data} provides a brief background of spatial computing and an overview of emerging spatial datasets at the cellular level. Section \ref{inner_space} discusses research needs across five broad areas of spatial computing research in inner space. Section \ref{applications_EHR} briefly discusses the potential of Atlas-EHR and provides examples across a wider range of medical fields. Finally, Section \ref{conclusion_FW} concludes the paper and outlines directions for future research.

\begin{table}[ht]\scriptsize
\centering
\caption{Taxonomy of Topics in This Paper}
\begin{tabular}{|c|c|c|c|c|c|}
\hline

\begin{tabular}[c]{@{}c@{}}Spatial Computing Topic\\ \\ \hline Study Area\end{tabular} & \begin{tabular}[c]{@{}c@{}}Spatial Database \\ Management\end{tabular}                          & \begin{tabular}[c]{@{}c@{}}Spatial Pattern \\ Mining (SPM)\end{tabular}                                                                                  & \begin{tabular}[c]{@{}c@{}}Positioning, Navigation, \\ Location-based services\end{tabular} & Sensing                                                                                  & \begin{tabular}[c]{@{}c@{}}Cartography \\ \&\\ Visualization\end{tabular} \\ \hline
Geographic Space 
    & 
    \begin{tabular}[c]{@{}c@{}}
    OGC's Simple feature types \cite{gardels1996open}, \\ R-trees \cite{gavrila1994r}
    \end{tabular} 
    & 
    \begin{tabular}[c]{@{}c@{}}
    Ring-shaped hotspots \cite{eftelioglu2014ring}, \\ 
    Local co-locations \cite{LocalLi2018}, \\ 
    CNN \cite{BengioLH21}, 
    GWR \cite{brunsdon1998geographically},  \\ 
    SVANN \cite{gupta2021spatial}\end{tabular} 
    
    & 
    GPS, WiFi 
    & \begin{tabular}[c]{@{}c@{}}
    RGB, Lidar, \\ 
    Multi-spectral, \\ 
    Hyper-spectral\end{tabular} & 
    GIS (e.g., ArcGIS) \\ \hline
    
Inner Space   
    &   
    & 
    \begin{tabular}[c]{@{}c@{}}
    CSCD \cite{li2022cscd}, 
    SRNet \cite{li2021srnet}, \\ 
    Spatially Explainable \\ Classifiers \cite{farhadloo2022samcnet}
    \end{tabular} 
    & 
    \begin{tabular}[c]{@{}c@{}}
    Talairach \\ coordinate system \cite{fischl1999high}
    \end{tabular}          
    & 
    \begin{tabular}[c]{@{}c@{}}
    X-rays, MRI, CT, \\ 
    MxIF \cite{lin2015highly}
    \end{tabular} 
    & 
    IonPath \cite{mccaffrey2020multiplexed}  \\ \hline
    
Outer Space  
    &  
    &      
    &     
    &    
    &    
    \\ \hline
\end{tabular}
\label{tab:taxanomy}
\end{table}



\section{Challenges and Opportunities: Technological Evolution in EHR and Spatial Computing.}\label{c&o}

Remarkable growth in electronic health records coupled with recent advances in data handling by AI and spatial computing make this an ideal time to pursue an Atlas-EHR initiative. The Patient Protection and Affordable Care Act (PPACA) of 2010, commonly known as Obamacare, has promoted the continuous development of EHRs \cite{ACA2024}, aiming to reduce healthcare costs and improve the quality of healthcare and clinical research. This includes facilitating recruitment for clinical trials, easing burdensome data collection, and addressing uncertainty in generalizing results, as detailed in \cite{cowie2017electronic}. However, EHR systems still encounter significant limitations and challenges. Key issues include interoperability, privacy, security, data ownership, governance, data quality and validation, complete data capture, and heterogeneity between systems, among others \cite{szarfman2022recommendations, holmes2021electronic, cowie2017electronic}. In recent years, there have been efforts to consider the role of emerging, big, complex, and heterogeneous EHR data in healthcare and its applications \cite{andreu2015big, kruse2016challenges, dash2019big, belle2015big, luo2016big, xie2022deep}. Yet, the full integration of this data across the spatial and temporal dimensions of the human body—an effort represented by Atlas-EHR—remains an emerging area of research and has not yet been fully realized.

For decades, designing pattern recognition and machine learning systems required deep engineering and domain-specific knowledge, particularly for creating effective feature extractors. These extractors convert raw data (e.g., image pixels), into feature vectors that the system’s classifier then uses for pattern recognition in the data. The rise of deep learning techniques and the training of neural network architectures with billions, and more recently, trillions of parameters (e.g., Google Switch Transformers \cite{fedus2022switch}), have significantly streamlined this intensive and computationally expensive process. This breakthrough has been made possible by the advent of fast graphics processing units (GPUs), along with the high performance computing centers, cloud computing, edge computing, and parallel computing techniques \cite{pal2019optimizing, dean2012large, wang2020convergence}. These GPUs, convenient to program, have enabled researchers to train networks 10 to 20 times faster \cite{lecun2015deep}. The application of deep learning in medical imaging is a prime example of this evolution. Machine learning models, such as convolutional neural networks (CNNs), have significantly improved the accuracy and speed of disease diagnosis from medical images \cite{shen2017deep, lundervold2019overview}. More recent techniques include the rise of graph convolutional neural networks (GCNNs) \cite{zhang2019graph, wu2020comprehensive}, which are well-suited for unstructured data, such as graph-based data commonly found in molecular structures, chemical compounds, and spatial pathology datasets (e.g., cellular maps). These datasets are critical for drug discovery \cite{chen2018rise}, protein-protein interaction networks \cite{jha2022prediction}, and gene-disease associations \cite{qumsiyeh2022gedinet}.

\textbf{Deep Neural Networks:} The success of convolutional neural networks (CNNs) in various applications, particularly in image recognition, is largely attributed to their ability to model spatial auto-correlation. CNNs prioritize spatially nearby samples over distant ones, capturing the inherent structure in images. This spatial awareness allows CNNs to outperform traditional fully-connected networks by a significant margin \cite{krizhevsky2012imagenet, lecun1998gradient}. However, a key challenge in handling spatial data, such as that found in EHRs, is spatial variability, that is the inherent heterogeneity and variations observed in spatial patterns, structures, properties, or arrangements across different locations \cite{gupta2021spatial}. The current deep learning literature \cite{guo2016deep} often adopts a spatial one-size-fits-all approach, training single deep neural networks that do not account for spatial variability. Preliminary works have aimed to address this challenge by considering location-dependent weights, learning space partitionings of heterogeneous data, and, more recently, through the use of meta-learning \cite{gupta2021spatial, xie2021statistically, liu2023task}. However, the applicability of these methods to EHRs and healthcare applications, given the multi-modal and multi-scale nature of electronic health record data remains largely unexplored.

\textbf{Foundation Models:} More recently, the AI community has witnessed significant breakthroughs in foundation models \cite{bommasani2021opportunities}, which can be fine-tuned for a broad range of tasks without needing task-specific data from scratch. Although a recent study \cite{mai2023opportunities} began to address the challenges associated with developing foundation models for geospatial AI, the potential of transferring these models from geospatial to biomedical scales remains largely unexplored. This gap offers a promising opportunity for the spatial computing community to innovate and broaden the application of foundation models to electronic health records. Investigating additional data modalities beyond images and text, along with integrating the temporal aspects of EHRs, are crucial open research areas. An equally paramount challenge lies in ensuring spatial explainability \cite{farhadloo2022samcnet}, which means the ability to describe results using spatial concepts (e.g., touch, inside)  and patterns (e.g., colocation). This is particularly relevant to oncologists seeking to advance new cancer therapies, as the spatial proximity between immune and cancer cells is crucial for evaluating the effectiveness of immune therapies. We further elaborate on this issue and explore related open research questions in Section \ref{inner_space}.

\textbf{Spatial Data Models:} Recent innovations in spatial database management systems, such as Spatial Hadoop \cite{eldawy2015spatialhadoop}, GeoMesa \cite{fox2013spatio}, and GeoSpark \cite{yu2019spatial}, have revolutionized data processing capabilities, making large-scale, complex spatial computations both feasible and efficient. These systems typically utilize established spatial data models, like Open Geospatial Consortium's (OGC) simple features (points, lines, polygons), to store, query, and manipulate large spatial datasets. However, challenges arise when modeling the human body and its organs, which are non-rigid and dynamic exhibits significant variability in size, shape, texture, and function. How do we define boundaries (e.g., minimum orthogonal boundary rectangles (MOBRs)) for spatial indexing techniques (e.g., R, R+ tress) given the non-rigid and soft shapes of the human body and its organs? Unlike geographic space, the human body lacks a reference system analogous to latitude/longitude, which is a challenge for spatial computing. Traditional models like the dimensionally extended 9-intersection model (DE-9IM) offer a framework for understanding spatial interactions (e.g., touch, inside). Yet, adapting these to account for hierarchical relationships and the dynamic changes associated with growth or disease that are time-dependent and subject to real-time changes remains a key challenge. For instance, the movement and deformation of body parts, such as muscles and joints, alter spatial relationships in intricate ways (e.g., parallel when the arm is extended, they form an acute angle upon bending). Additionally, fluid movement within body cavities, like blood in vessels or air in respiratory pathways, involves dynamic interactions that challenge static spatial modeling. Certain relationships, like bacteria within the gut or within a cell, are "inside" not just in terms of spatial positioning but also in terms of their functional roles and dependencies. Addressing these issues presents the spatial computing community with unique opportunities to explore.

\section{Background \& Emerging Spatial Data from Inner Space }\label{emergin_spatial_data}

\subsection{Brief Overview of Spatial Computing}\label{spatial_computing_overview}
Spatial Computing \cite{shekhar2015spatial, shekhar2020spatial, eftelioglu2017nexus} encompasses  a diverse array of tools, frameworks, solutions, and technologies, collectively transforming our understanding and interaction with spatial data. This transformation extends beyond simple navigation; it influences how we perceive, communicate, and visualize relationships with various locations. A prime example is the Global Positioning System (GPS), a critical infrastructure for the world economy considering over 2 billion receivers in use for location and time services, with widespread applications from COVID-19 response to smart city development.

One of the main areas of spatial computing research is cartography and geo-visualization \cite{crampton2018introduction, maceachren2013visualization, maceachren2004maps, maceachren1997exploratory}. This field extends beyond conventional data visualization by incorporating spatiotemporal elements and enables the representation and analysis of data to its physical or spatial location over time.  Geographic information systems, such as ArcGIS \cite{esriarcgis2024} and QGIS \cite{QGIS2024}, are widely used applications in diverse domains, such as public health, public safety, and agriculture. These systems offer a user-friendly interface to import data from third-party sources (e.g., the US Census), utilize spatial statistics techniques (e.g., inverse distance weighting, Kriging) \cite{scott2009spatial, johnston2001using} for interpolation, and to visualize derived patterns (e.g., air quality interpolation) in maps.

\begin{figure}[h!]
     \centering
     \begin{subfigure}[b]{0.40\linewidth}
         \centering
         \includegraphics[width=1.0\linewidth, height=8cm]{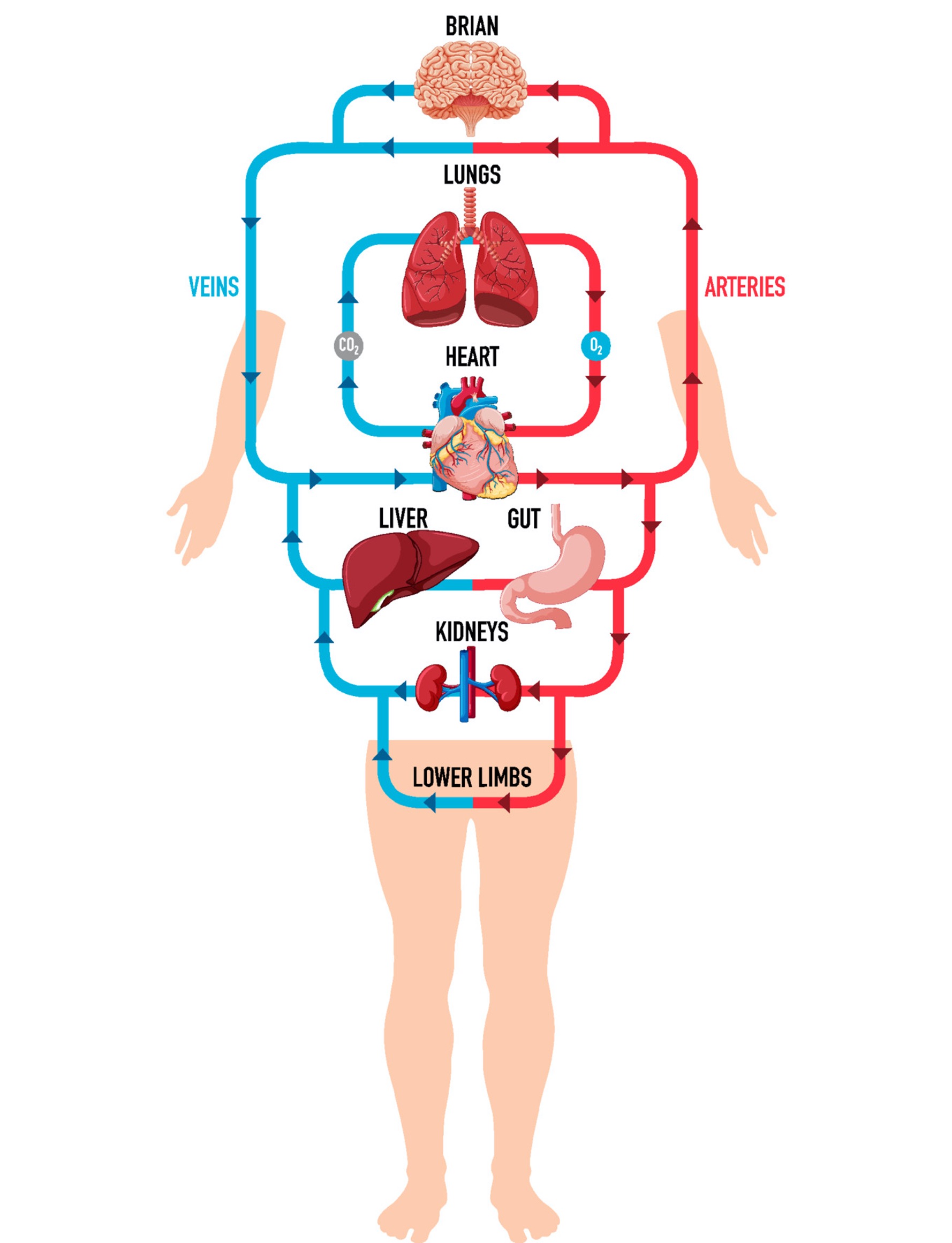}
         \caption{A transit network of arteries (red) and veins (blue) for blood transport in the human body.}
         \label{human_map_a}
     \end{subfigure}
     \hfill
     \begin{subfigure}[b]{0.40\linewidth}
         \centering
         \includegraphics[width=.85\linewidth, height=8cm]{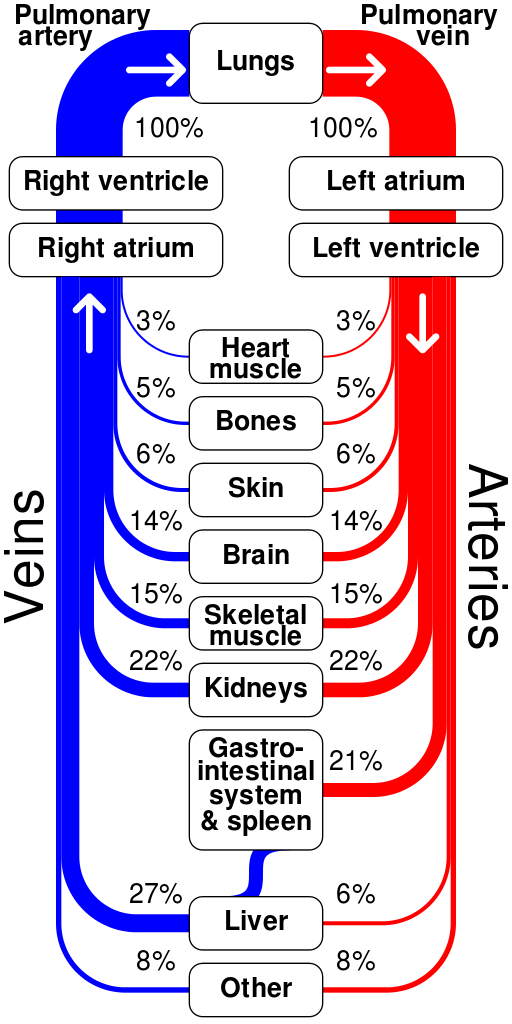}
         \caption{Sankey diagram representing the human circulatory system, showing the distribution of cardiac output to various organs through arteries (red) and veins (blue).}
         \label{human_map_b}
     \end{subfigure}
     \caption{Spatial maps of the human circulatory system, illustrating the flow and distribution of blood to various organs.}
\end{figure}

From the Babylonian world map circa 600 B.C. \cite{horowitz1988babylonian} through the pioneering geographical works of Eratosthenes, the father of geography, to the development of the Mercator projection \cite{snyder1978space} for sailing and global navigation, and the mapping of rail transport networks \cite{siebert2004using}, cartography and geo-visualization have been pivotal in human innovation for centuries. They enable exploration and navigation across the globe. The rich history of cartography, with its emphasis on geographical accuracy and detail, combined with modern principles of data visualization \cite{gandhi2020data,wanderer2016clinical, few2007data} and human-computer interaction (HCI) \cite{mackenzie2024human, carroll1997human, myers1998brief, hochheiser2020human}, opens new avenues for research. By leveraging cartographic insights for spatial orientation and accuracy, alongside data visualization for abstract data representation, and HCI for intuitive user engagement, we have the means to create more comprehensive and accessible visualizations to explore the human body in our Atlas-EHR vision. 
Figure \ref{human_map_a} and \ref{human_map_b} shows two simple illustrations. Figure \ref{human_map_a} depicts the human circulatory system, mapping veins and arteries critical for blood transport—a system akin to transit stops, essential for oxygen and nutrient delivery and waste removal. \ref{human_map_b}, is a sankey map of the circulatory system, depicting the flow of oxygenated blood (similar to Minard's flow map of Napoleon's invasion of Russia \cite{nationalGeographic2024}) showcasing the integration of statistical graphics into a human anatomy map. More details on challenges and opportunities will be discussed in Section \ref{inner_space}.

\subsection{Spatial Data at the Scale of Inner Space}
Emerging spatial omics technologies are providing researchers with biomedical data at the cellular and molecular level. These technologies analyze RNA molecules within cells. RNA acts as a messenger that takes the genetic instructions from DNA and uses them to produce proteins that the cell needs. The types and amounts of proteins created depend on the cell's condition. For example, a healthy cell will produce different proteins compared to a diseased cell. By studying RNA, spatial omics reveals which proteins a cell is making under different physiological or pathological states. This gives key insights into cellular and tissue function that are essential for health and survival.



The Human Cell Atlas (HCA) \cite{rozenblatt2017human} is a global collaborative effort aimed at creating a detailed map of every human cell type to fundamentally enhance our understanding of health and disease mechanisms. Initiated by researchers around the world, the HCA seeks to characterize cells not only genetically and functionally but also in terms of their spatial distribution within tissues. This ambitious project endeavors to transform our approach to diagnosing and treating diseases by providing an unprecedented level of insight into the cellular basis of human biology. Through its comprehensive cellular mapping, the HCA promises to lay the groundwork for significant advancements in medical research and healthcare and offers a more nuanced understanding of the human body's inner workings.


The Human Tumor Atlas Network (HTAN) \cite{HTAN2024, wu2021single, uhlen2017pathology, rozenblatt2020human} focuses on developing comprehensive atlases of tumor development to understand cancer evolution through the integration of spatial and temporal data. By mapping the progression of various cancers in detail, HTAN enhances our understanding of cancer biology, which in turn facilitates the creation of more effective therapies. This network represents a concerted effort to dissect the complexities of cancer, leveraging advanced technologies to capture the dynamic changes in tumors over time and across different environments, ultimately contributing to improved cancer diagnosis, treatment, and management strategies.

The Human Cell Atlas (HCA) and Human Tumor Atlas Network (HTAN) represent significant strides in biomedical research, yet they primarily concentrate on gathering data and providing tools for basic cell interactions. Studies of more intricate spatial cellular interactions and variations in individual cell composition remain largely unexplored and present significant opportunities for further development within spatial computing research. Moreover, the Human BioMolecular Atlas Program (HuBMAP)\cite{hubmap2019human} aims to develop detailed maps of human biomolecules at the cellular level. A notable achievement of HuBMAP is the creation of Organ Mapping Antibody Panels (OMAPs) for complex spatial imaging in seven human organs. The program also provide computational tools like STELLAR for annotating single-cell data and segmenting specific structures within tissues. The Cellular Senescence Network (SenNet) Program, initiated by the Common Fund, aims to conduct a thorough investigation and differentiation of senescent cells within the human body, across various health conditions, and throughout the human lifespan. This endeavor seeks to map and understand the complex nature of cellular senescence in relation to human health and aging. SenNet leverages advancements in single-cell analysis, similar to those in HuBMAP, but with a specific emphasis on senescent cells, providing tools, technologies, and a publicly accessible atlas of these cells and their secretions. However, oncologists are interested in understanding the spatial interactions among immune and cancer cells in situ within the tumor microenvironment (TME) in order to gain insights into the mechanism of immune therapies and inform the development of new and more effective treatments.

Developing research in this field highlights the need for an automated process to analyze the complex spatial relationships among different cellular subsets and functional states within the tumor microenvironment (TME) in the context of immune checkpoint inhibitor (ICI) therapy\cite{maus2022resolving, hoch2022multiplexed}. Furthermore, it is clinically crucial to examine the importance of cell species along with their functional status in a spatially informed manner, due to the clinical implications of interactions in close spatial proximity. Analyzing such spatial interactions helps generate new hypotheses towards discovering disease therapeutics (e.g., immunotherapies for cancer treatment) and could be used in applications such as medical pathology, biomedical research, and microbial ecology. This gives rise to multiple challenges, such as multi-category point patterns, which are heterogeneous and form complicated structural and higher-order spatial interactions. In addition, this can also result in an exponential number of category subsets, which may vary in the strength of their spatial interactions. The proposed Atlas-EHR envisions handling such spatial heterogeneity for both global and local frames of reference and further provides GPU capability to speed up analysis of exponential multi-categorical relationships while preserving spatial interpretability for oncologists.

\section{Accomplishments and Open Research Opportunities in Inner Space}\label{inner_space}
This section lists a few challenges and open research questions related to spatial computing in inner space into five broad areas. It highlights specific examples and challenges within each sub-area, including the development of new spatial data models to describe the intricate spatial relationships within the human body, addressing privacy and ethical concerns during data collection and processing, generating and analyzing large-scale cellular maps, creating spatially-explainable models, managing spatial and patient variability, mitigating spatial bias and ensuring fairness, and developing spatially-enabled generative AI techniques for visualizing cellular maps and simulating biological behaviors.


\subsection{Spatial Database Management systems} 
The fundamental inquiry focuses on the interactions among spatial entities (e.g., vital organs, their components, subcomponents, tissues, cells, etc.) within the inner space, which depends on an underlying spatial framework defining their relationships. To make this concept more relatable, imagine the human body as a city, where organs represent various departments; for instance, the heart functions as the water pump station, and the lungs act as the air filtration system. These `departments' interact and communicate to maintain the health and efficiency of the city or body. Take the gut-brain axis \cite{carabotti2015gut} as an example: it serves as the communication line between the city's food supply chain (gut) and the city management center (brain), which controls the city's mood (emotions) and operational efficiency (digestion and nutrient absorption). Similarly, the heart acts as the central pump in the water system and the lungs as a purification plant, where air entering the lungs is akin to water that is being filtered. The heart then pumps this `cleaned' air (oxygen-rich blood) through the body's `pipeline system' to ensure every part receives the `freshwater' necessary for proper function. Spatial database management systems act as the city's mapping and information hub, organizing data about the interactions between departments and aiding in the understanding and optimization of the body's functions.

\textbf{Spatial Data Models.} 
While OGC's simple features (e.g., multi-points, multi-lines, multi-polygons) \cite{gardels1996open} offer a solid foundation for representing common shapes on a map through a spatial reference system, accurately representing and analyzing spatial relationships in inner space can be challenging. These include the non-rigid and flexible nature of soft tissues and organs, their dynamic changes, ability to compress and expand, pathological changes that alter spatial boundaries, the existence of multi-scale structures, the distortion from projecting three-dimensional organs onto two-dimensional maps, and the lack of a spatial reference system tailored to the human body. Addressing these issues requires advancing beyond traditional data models and operations designed for geographic spatial entities.

To illustrate, topological relationships (e.g., disjoint, meet, overlap) highlight properties that remain constant under continuous transformations, whereas non-topological relationships (e.g., metric relationships such as distance and angle) can change with transformations. These are typically defined through the existing 9-Intersection Model and the queen-based neighborhood directional model. However, capturing the intricate spatial relationships within the human body requires considering a combination of these relationships, extending beyond pairs to include non-distance-based spatial associations (e.g., \textit{surrounded by}) that are invariant to elastic deformation.

For instance, understanding the relationship between the heart and lungs in the cardiovascular and respiratory systems involves recognizing how oxygen-poor blood is pumped from the heart to the lungs (a topological relationship), where it undergoes oxygenation (a non-topological process involving gas exchange), and is then returned to the heart for systemic circulation. Therefore, spatial data models must incorporate anatomical pathways (veins and arteries) and physiological processes (oxygenation levels, blood flow rates) to examine a complex interplay between structure (topology) and function (non-topology) across vital organs. Similarly, the interaction between the digestive and circulatory systems involves the liver's role in processing nutrients from the intestine and its complex vascular connections to other organs. Blood from the digestive tract is transported to the liver via the portal vein (topological), where it is filtered and metabolized (non-topological, involving chemical processes). The liver then releases nutrients into the bloodstream for distribution throughout the body, exemplifying the integration of spatial pathways with biochemical transformations across different organs.


\textbf{Big Spatial Data at Inner Space Scale.} 
Current studies \cite{berry2021analysis, bressan2023dawn, lewis2021spatial, velten2023principles, yuan2023sodb}  in spatial omics and cellular mapping often analyze a limited number of pathology-driven fields of view (FOVs), typically ranging from five to ten, to investigate spatial cellular interactions within the tumor microenvironment (i.e., one slide). However, comprehensively understanding the tumor microenvironment requires analyzing hundreds of FOVs, potentially reaching 300 for a detailed picture. If the aim is to capture all tissue regions (normal, boundary, malignant), the number of FOVs can exceed 1300. A recent study \cite{berry2021analysis} has captured entire slide images, generating massive datasets reaching five terabytes.
Considering that a single tumor biopsy is typically divided into several slides for targeted analysis, mapping an entire biopsy at high resolution could require up to 100 slides. If the tumor affects multiple organs and involves numerous patients with similar conditions, the data volume could easily scale to hundreds of petabytes.

This big spatial omics data is essential because the spatial relationships and patterns within it are crucial for understanding biological processes and cannot be fully captured without comprehensive and integrated analysis methods. While parallel processing and time-staggered analysis can distribute the computational load, they do not address the fundamental challenges of data storage, retrieval, and real-time analysis demands. Distributed systems and parallel processing across different machines or sequential analysis over time can mitigate some computational burdens, but they do not directly solve the challenge of integrating diverse data types (e.g., imaging, genomic, proteomic) to derive meaningful insights. Additional complexities arise due to variability in biological processing across different organs (i.e., different locations within the body) and their associated subcomponents, which can vary significantly and pose significant challenges for existing spatial big data management systems. For instance, diseases such as  Alzheimer's or Parkinson's necessitate time-critical analyses, requiring real-time processing of spatial data to monitor physiological changes in brain over time.

This demonstrates the potential and need to design new spatial data models and modify existing spatial indexing, query processing, and optimization for effectively handling massive spatial data at scale. For example, is it necessary to capture the whole slide image? To reduce the expense of storing hundreds of petabytes of data, would it be possible to consider a range query or a nearest neighbor algorithm to estimate the spatial distribution of the whole slide image, perhaps followed by imposing a non-uniform spatial grid index to sample FOVs? With an estimate of the spatial distribution of the whole slide image, is it possible to define the cost and route over a spatial network of FOVs to reach areas associated with the high proliferation of tumor cells?  

\begin{figure}[ht]
    \centering
    \includegraphics[width=0.5\textwidth]{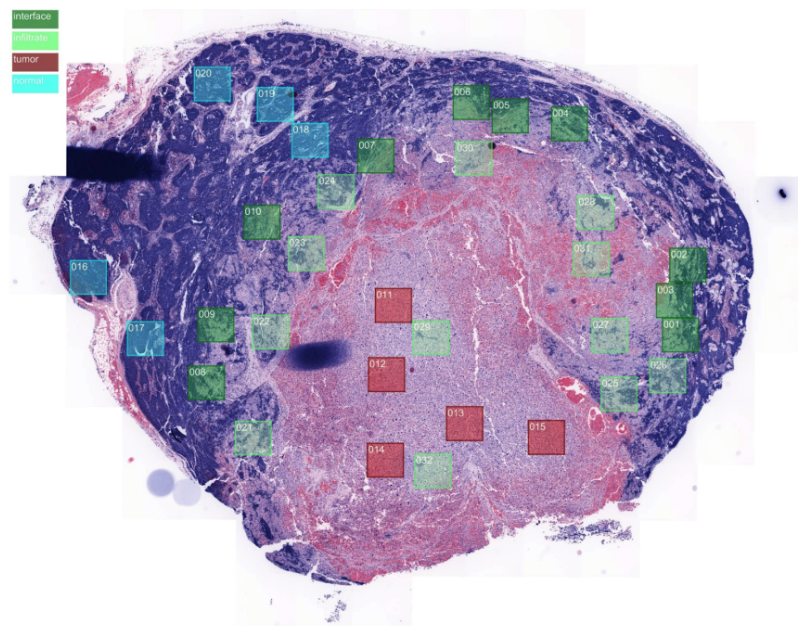}
    \caption{A virtual  H\&E tissue slide including pathology-driven fields of view namely, infiltrate, interface, tumor, and normal \cite{maus2022resolving}.}
    \label{fig:mxif_tissue}
\end{figure}

Figure \ref{fig:mxif_tissue} shows pathology-driven FOVs from a virtual H\&E tissue slide, which includes tumor, infiltrate, interface, and normal FOVs. The current inspection process is expensive and time-intensive due to the need for domain experts (e.g., pathologists) to manually analyze tissue slides and label them according to response to antibody staining. The current solution is to capture the whole slide image by imposing an overlapping grid structure; however, the associated data storage cost is high. We envision an algorithm that begins with a stochastic approach to generate random FOVs over the whole slide image. This is followed by a spatial network, where nodes are individual FOVs and edges describe the similarity between FOVs based on a summary statistic to estimate the spatial distribution at the FOV-level. This process can be repeated until an optimal spatial distribution is reached, meaning when adequate FOV group samples are generated with high intra-similarity and minimum inter-similarity between groups. However, additional complexities may arise due to the impact of the edge effect and signal-to-noise ratio in estimating spatial distribution.  

\textbf{Spatial Physical Design for Omics ML}. 
The following research questions highlight the lack of spatial components in recent machine learning algorithms, including multi-task and federated learning. Is it possible to model the underlying spatial distribution of sampled FOVs from specific tissues (e.g., skin, kidney, lungs) to define spatial indices over regions that share similar spatial configurations? If successful, this may result in millions to billions of parameters of pre-trained models related to specific tasks on big spatial omics data. What is the best way to define a standard spatial projection and reference system to leverage large data samples with different formats across multiple sources? Do pre-trained models need to be stored on disk or memory, or do we need to define intermediate stages on the database that save some parameters that can then be fine-tuned for other specific tasks? How can spatial omics data be most effectively prepared at scale to be used by machine learning models on related downstream tasks?

Multi-task learning aims to improve the performance of individual tasks, where a model is trained on multiple tasks simultaneously by leveraging the shared representations learned across all of them. Federated learning has also gained attention in past years. Federated learning is a distributed ML technique where a model is trained across multiple decentralized devices without sharing the raw data with a central server. This becomes important since patient privacy and the propriety nature of medical data prevent many organizations (e.g., hospitals and medical research establishments) from publishing their datasets. Federated learning is an effort to address this issue by only transferring model parameters while training the model on locally stored data. Such algorithms, however, may not be generalizable across all domains since recent works \cite{palla2022spatial, lewis2021spatial} have highlighted that analyzing spatial omics data is location-specific within the body (e.g., brain, kidney, lungs), and separate training models are needed to better understand the structure, composition, and underlying spatial patterns. This demonstrates the critical role of spatial database management and relevant techniques (e.g., spatial indexing) for preparing these data at scale to be used by machine learning models on related downstream tasks (e.g., classification, segmentation).

\subsection{Spatial Pattern Mining} 

\textbf{Spatially-Explainable Prediction Models.} 
Domain scientists (e.g., oncologists and pathologists) are interested in understanding why patients respond differently to a candidate immunotherapies. Success hinges on complex spatial interactions between cancer and immune cells within tumor microenvironments. Studying these manually is increasingly challenging due to the exploding volume and richness of spatial omics and cellular maps generated by modern imaging. For example, a single tissue sample contains millions of cells representing over 60 distinct cell sub-types, which results in a potential million trillion spatial cell interactions to investigate. In addition, the strength of these interactions can vary, with some interactions being more biologically relevant than others, requiring the model to represent these distinct levels of interaction. 

A possible research direction is using advanced classification models (e.g., adaboost, xgboost, KNN, SVM ) and state-of-the-art deep neural networks (DNNs) for downstream tasks (e.g., predictions, classifications). However, these models do not provide spatial explainability. Hence, is it possible to describe the clinical outcomes of an intervention using spatial concepts (e.g., \textit{touch, inside, surrounded by}) and patterns (e.g., colocations, spatial outlier, hotspot-colocations)? The early spatial approaches used hand-crafted measures from classical spatial statistics (e.g., Ripley’s cross-k function, G-cross) \cite{dixon2001ripley, barua2018spatial} and spatial data mining (e.g., participation index) \cite{huang2004discovering, shekhar2001discovering, yoo2006joinless, li2022cscd}, are primarily based on distance relationships. These do not model directional spatial relationships such as \textit{surrounded by}, which may be of biological significance. 

Explainable artificial intelligence (XAI) \cite{confalonieri2021historical, dovsilovic2018explainable, angelov2021explainable}  encompasses algorithms designed to enhance transparency and trust by providing understandable explanations for their decisions while maintaining acceptable performance. This becomes even more crucial for spatially-explainable AI approaches when analyzing spatial patterns in multi-category point sets like spatial omics and cellular maps. This research holds significant importance in advancing immune checkpoint inhibitor (ICI) cancer therapies, where the effectiveness hinges on spatial proximity between immune and cancer cells triggering crucial biochemical interactions. Our recent works \cite{li2021srnet, farhadloo2022samcnet} has shown promise for machine-constructed spatial cell interactions in modeling non-distance-based spatial relationships, outperforming state-of-the-art methods in separating responder from non-responder classes. This work is a step toward a  spatially explainable AI-based approach, which involves classifying data into a given number of classes and determining the most discriminative features based on their spatial arrangement. This is only the beginning, as many other opportunities remain, including the development of spatially explainable AI methods for other common pattern families, such as spatial hotspots and anomalies. In addition, is it possible to detect even more complex spatial interactions using spatially explainable data mining, not only in cellular maps in oncology but maps found in other domains?

\begin{figure}[ht]
    \centering
    \includegraphics[width=0.8\textwidth]{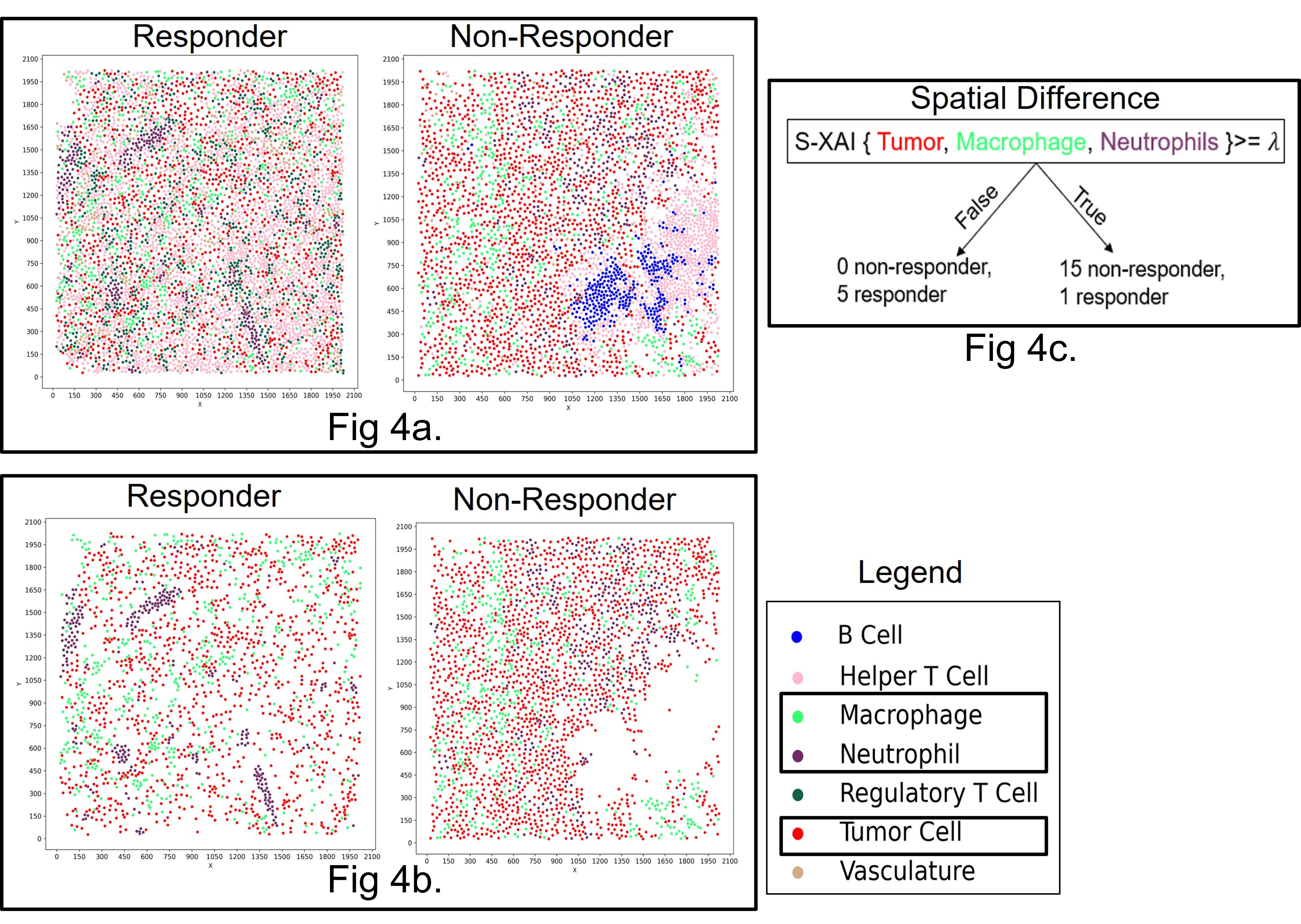}
    \caption{(a) Multi-category point sets of the responder and non-responder (b) Point patterns detailing spatial arrangements between Tumor cells, Neutrophils, and Macrophages. (c) A spatially-explainable decision tree to distinguish between responder and non-responder at the tumor core}
    \label{fig:prelim}
\end{figure}

Our preliminary work \cite{farhadloo2022samcnet} sought to identify the spatial differences that distinguish  responders and non-responders at the tumor core. A novel neural network architecture was designed for the cellular map data to discover any significant spatially-explainable patterns that could not be captured by traditional spatial data mining or state-of-the-art DNN techniques. This framework introduces two new layers in the neural network for classifying cell interactions. The first layer models cellular maps belonging to different distributions (e.g., clustering versus even distribution), and the second distinguishes between cell-cell instances that belong to different types by learning the importance of each distinctive pair. Figure \ref{fig:prelim} (a) shows the multi-category point sets for all cell species in a responder class compared to a non-responder class. The significance of the identified cell interactions, namely, the machine-constructed spatial cell interactions for different category subsets, was evaluated by feature permutation importance. The top important feature was identified as spatial cell interactions among <tumor, macrophage, and neutrophils>, where macrophage is a type of immune cell, and neutrophil is a type of immune cell. This analysis revealed (Figure \ref{fig:prelim} (b)) that macrophages were associated with tumor cells in both the responder and non-responder classes. It also showed that neutrophils formed small sub-clusters in the responder tissue sample, while in the non-responder tissue sample, they surrounded the tumor-associated macrophage. A spatially-explainable AI classifier (Figure \ref{fig:prelim} (c)) uses the spatial differences between these cells as explanatory features to distinguish between responder and non-responder multi-category point sets.

\begin{figure}[ht]
    \centering
    \includegraphics[width=0.7\textwidth]{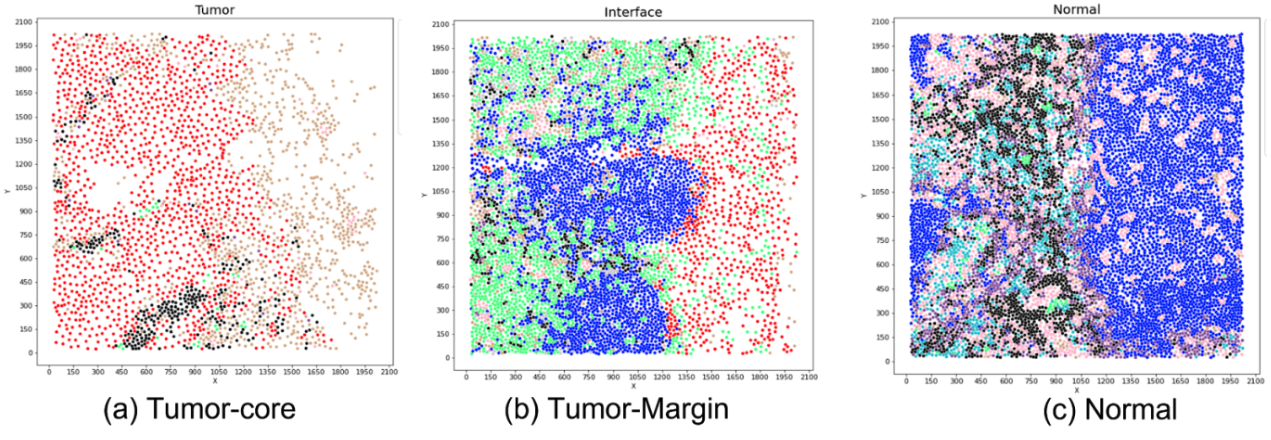}
    \caption{Three pathology-driven fields of view are categorized based on tumor infiltration} 
    \label{fig:spatial_variabilities}
\end{figure}

\textbf{Spatial Variability Aware AI Models.} 
Spatial variability introduces additional complexities to the analysis of multi-category point sets. Spatial variability refers to the inherent heterogeneity and variations observed in spatial patterns, structures, or arrangements across different locations. For example, in Fig. \ref{fig:spatial_variabilities}, the three selected multi-category point sets (from Fig. \ref{fig:mxif_tissue}) depicting place-types (e.g., tumor-core, tumor-margin, and normal regions) illustrate significant spatial variability, revealing variations in cell population across different locations within a single tissue sample. Beyond variability across place-types, intricacies emerge within their sub-categories. For instance, tumor margins can be classified as intact or infiltrated based on tumor boundary properties. Intact regions, displaying negative response, have boundaries between immune and tumor cells remaining intact. Infiltrated regions, on the other hand, can represent either a favorable response (immune cells invading and destroying cancer) or a negative response (tumor cells attacking immune cells). Spatial variability can further extend to sub-regions within a place-type, where specific patterns like hotspots might be more prevalent than others, such as co-locations and anomalies.

While recent studies \cite{nawaz2016computational, yuan2016spatial}, offers insight into tumor microenvironment heterogeneity, modeling its spatial variability is challenging due to factors such as rapid cancer cell proliferation, genetic instability, and the presence of unknown mediator cells with immune and target cells. These complexities demonstrate the need for spatial variability aware pattern mining in analyzing cellular maps in inner space. 

Geographically Weighted Regression (GWR) \cite{brunsdon1999some} is a traditional non-parametric model that learns location-specific maps but struggles with complex predictions. Recent work such as SVANN \cite{gupta2021spatial} and \cite{xie2021statistically} study spatial variability through deep learning considering location-dependent weights and learning space partitionings of heterogeneous data using weight-sharing mechanisms. However, techniques require dense training data or have limited ability to handle significant spatial variability within a single place-type. Moreover, they are designed for geographic data in Euclidean space and miss the complexities of spatial variability in non-Euclidean spaces like point clouds and multi-category point sets. The advancement of CNNs \cite{lecun2015deep, he2016deep, cecotti2020grape} in pattern recognition has inspired their adaptation for analyzing 2D/3D point cloud data without the need for expensive conversion steps. Pioneering deep neural networks like PointNet \cite{qi2017pointnet}, DGCNN \cite{wang2019dynamic}, and the more recent PointTransformers \cite{zhao2021point} have significantly improved the processing of point cloud data for pattern recognition. However, these methods primarily focus on numerical attributes, which limits their ability to capture the complex spatial relationships among various categories of points. SAMCNet \cite{farhadloo2022samcnet} addresses this gap by emphasizing spatial interactions across diverse point types but does not adequately account for \textit{spatial variability}, with a preference for scalar over map weights. 

Our recent work \cite{farhadlooSiam2024} explores a spatial ensemble framework, learning unique decision functions for different place-types (e.g., tumor-core, tumor-margin), and hypothesizes that a spatial ensemble of base classifiers would enhance prediction strength and model interpretability. The network parameter is a map varying across place-types rather than a scalar as used in traditional size-fits-all approaches. Additionally, this work investigates more flexible spatial domain adaptation layers to tackle issues with insufficient learning samples, using a weight-sharing mechanism for location independent and dependent layers. Location-independent layers are learned using all cellular maps across all regions, while subsequent layers are fine-tuned to capture location-dependent features for the target domain classifier. This approach mirrors the innovation seen in foundation models \cite{bommasani2021opportunities}, where it's envisioned that location-independent features (domain-independent) in a target domain classifier can be derived from the original foundation model without the need for task-specific data from scratch. However, location-specific features that differ across place-types still necessitate fine-tuning. Recent advancements include the use of generative adversarial networks \cite{creswell2018generative} for domain adaptation \cite{wang2018deep, zhuang2020comprehensive} that leverages minmax optimization  to train domain-specific discriminators and a generative model that seeks to produce domain-invariant features.  The application of these techniques to spatial datasets has yet to be fully explored \cite{mai2022towards, mai2023opportunities}.


\subsection{Positioning, Navigation, Location-based services} 
Computer-vision positioning \cite{tong2023multi} is a technique that aims to determine the position and orientation of objects by analyzing visual data, such as images.  This technique also can provide coordinates and descriptive text corresponding to the scene being analyzed. However,  directly applying these techniques to inner space may not work correctly due to patient variability (i.e., patient-specific) and the non-rigid structure of human organs, muscles, blood vessels, etc. For instance, the size and position of the liver can vary significantly due to factors like age, gender, and the presence of certain diseases. Patient variability could be classified based on a hierarchy of demographic, health, and physical fitness information. From a high-level view, patients may be classified by their age, race, ethnicity, and gender, factors that have a high impact on the inner space of human physiology. Additional complexities may arise due to the patient's medical history (e.g., past injuries, chronic illness, etc.) and physical fitness. At this level, computer-vision-positioning faces a great challenge due to the data heterogeneity at multiple phases. Even within a single individual, organs and their sub-components can exhibit significant variability in morphology. This spatial variability adds another layer of complexity for positioning and analysis.   

A possible solution may be defining a heterogenous data-aware hierarchy to split the data to account for patient variability. Such a hierarchy could be specified as a procedure to characterize a social/fair/mathematical norm such that in-depth features (e.g., detailed physical or health history) impact how the dataset is partitioned.  Beyond patient variability, the non-rigid structure of human organs could highly impact how computer-vision-positioning technology should be modified and adapted to inner space. Human organs and muscles, including the heart, kidneys, liver, and cardiac and smooth muscles, can change shape and size in response to various physiological and pathological processes, and the frequency of these changes can vary widely depending on the specific organ/muscle and the conditions it is subjected to. For example, the heart expands, and contracts with each beat, and the size of the ventricles can change in response to changes in fluid volume. The lungs can expand and contract with each breath, and the size of the air spaces can change in breathing rate or lung volume. This elastic deformation of human organs with high temporal variability calls into question the generalizability of computer-vision-positioning. Is it possible to train a learning algorithm on images of a certain physical structure of some organs and use those images to determine the elastic deformation of the same organ at the next temporal step? 

\subsection{Sensing} 
Recent advancements in AI, machine learning, sensor miniaturization, and battery life have revolutionized smart wearable devices. These devices have become more comfortable, less intrusive, and capable of monitoring a broader range of health metrics \cite{sabry2022machine, shajari2023emergence}. They enable continuous and real-time data collection that is driven by improvements in the Internet of Things (IoT) \cite{rose2015internet, li2015internet, qadri2020future, aghdam2021role}. For example, devices that monitor and rehabilitate gait in people with Parkinson’s disease (PD) exemplify the practical applications of these technologies. These smart devices, ranging from smartwatches to body-worn sensors, are designed to provide objective, quantitative data on the motor symptoms of PD, such as slowness of movement and involuntary muscle movements \cite{moreau2023overview, kubota2016machine}. Another innovative development is smart clothing, which excels particularly in cardiovascular risk assessment and monitoring by leveraging sensors like electrocardiograms (ECGs) and photoplethysmography (PPGs). These garments provide continuous measurement of vital health indicators like heart rate variability and microvascular blood volume changes, which are crucial for predicting cardiovascular risks \cite{bayoumy2021smart, moshawrab2023smart}. However, several challenges still hinder the widespread adoption of wearables in clinical practice, including concerns about device accuracy, patient privacy, and cost, and the difficulty of separating actionable data from noise. Additionally, well-designed trials are needed to fully establish their advantages, all of which are detailed in recent studies  \cite{bayoumy2021smart, piwek2016rise, steinhubl2015emerging, tison2018passive}.

Beyond IoT-enabled wearable and smart devices, medical imaging is a vital technique used in healthcare. It falls within the broader definition of remote sensing where electromagnetic radiation or sound waves are emitted towards the body and their interactions provide information for visualizing human internal structures and organs. While these technologies could provide high-throughput spatial information at a great details across different modalities (e.g., MRI, CT scans) and scales (e.g., whole-body, organ, cell, molecular, genetic), they are limited to providing snapshots and cannot continuously monitor dynamic physiological processes (e.g., blood flow and cardiac function) in real-time. In addition, current medical diagnostics involve the extraction of a sampling tissue, which is difficult to do in sensitive areas, such as the brain. Can we generate some minimally invasive methods for such sensitive areas which can provide more value in early-stage diagnosis? 

\textbf{Towards a Hybrid Future: Merging Wearables and Imaging.} While smart wearables provide continuous health data, they have accuracy limitations. Medical imaging gives high-resolution spatial snapshots but lacks dynamic physiological information. Could a hybrid approach that combines sparse wearable data streams with biophysics-based modeling reconstruct a 4D volumetric processes? Examples include real-time cardiovascular dynamics monitoring that requires detailed 4D models of blood flow and cardiac dynamics. It it possible to capture this through integrating sparse, noisy data from wearable sensors with spatially and temporally coherent biophysical models?  Accounting for missing data, modality-specific noise and artifacts, aligning different coordinate systems, fusing data collected at different frequencies, and privacy-preserving analytics provide challeneges for current spatial sensing data pipelines. Developing biocompatible sensors that can provide high spatiotemporal resolution signals from sensitive organs without negatively impacting surrounding organs and nervese system pose additional complexities. 

Augmented reality (AR) can improve surgical precision, speed, and safety by providing live visual overlays like critical anatomy in the surgeon's field of view. This approach can reduce complications and accelerate the learning curve for surgeons. Challenges include ensuring accurate tracking and registration of overlays, robust occlusion handling, developing prototype systems that integrate smoothly into surgical workflows, user-centric spatial interaction techniques, and conducting user studies to quantify their value.

\subsection{Cartography and Visualization} 
Atlas-EHR, an inner space information system, serves as an integrated platform that includes the previously discussed techniques to tackle the outlined challenges.
This unified framework is adept at collecting, managing, analyzing, and visualizing EHR data. Specifically, it incorporates spatially-explainable AI approaches tailored for various common pattern types. These include spatial co-location among different cell subsets, identification of cancerous and metastasized cell hotspots, and mapping the telecommunication pathways from the brain to vital organs. The system can also integrate these patterns, considering the complex nature of EHR data. 

A key principle for understanding the significance of problems, common patterns, and relationships in data lies in exploratory analysis and visualization before employing advanced computational techniques, such as machine learning. In cancer research, understanding certain spatial cell interactions requires the selection of contrasting colors, sizes, and other visualization components to emphasize specific spatial patterns. However, existing GIS software has limited capabilities for automating the visualization of cells of different classes. This results in extensive manual effort. Considering the increasing number of identified cell types, their functional characteristics, and the thousands of samples derived from specific tissues, the cost of manual visualization can increase exponentially. The question arises whether we can enhance existing GIS software capabilities through the innovation of generative AI algorithms suitable for inner space exploration. More specifically, is it feasible to implement a spatially-aware generative AI model that enables end-users to easily visualize cellular maps and distinguish specific spatial patterns with minimal commands?

\begin{figure}[ht]
    \centering
    \includegraphics[width=0.5\textwidth]{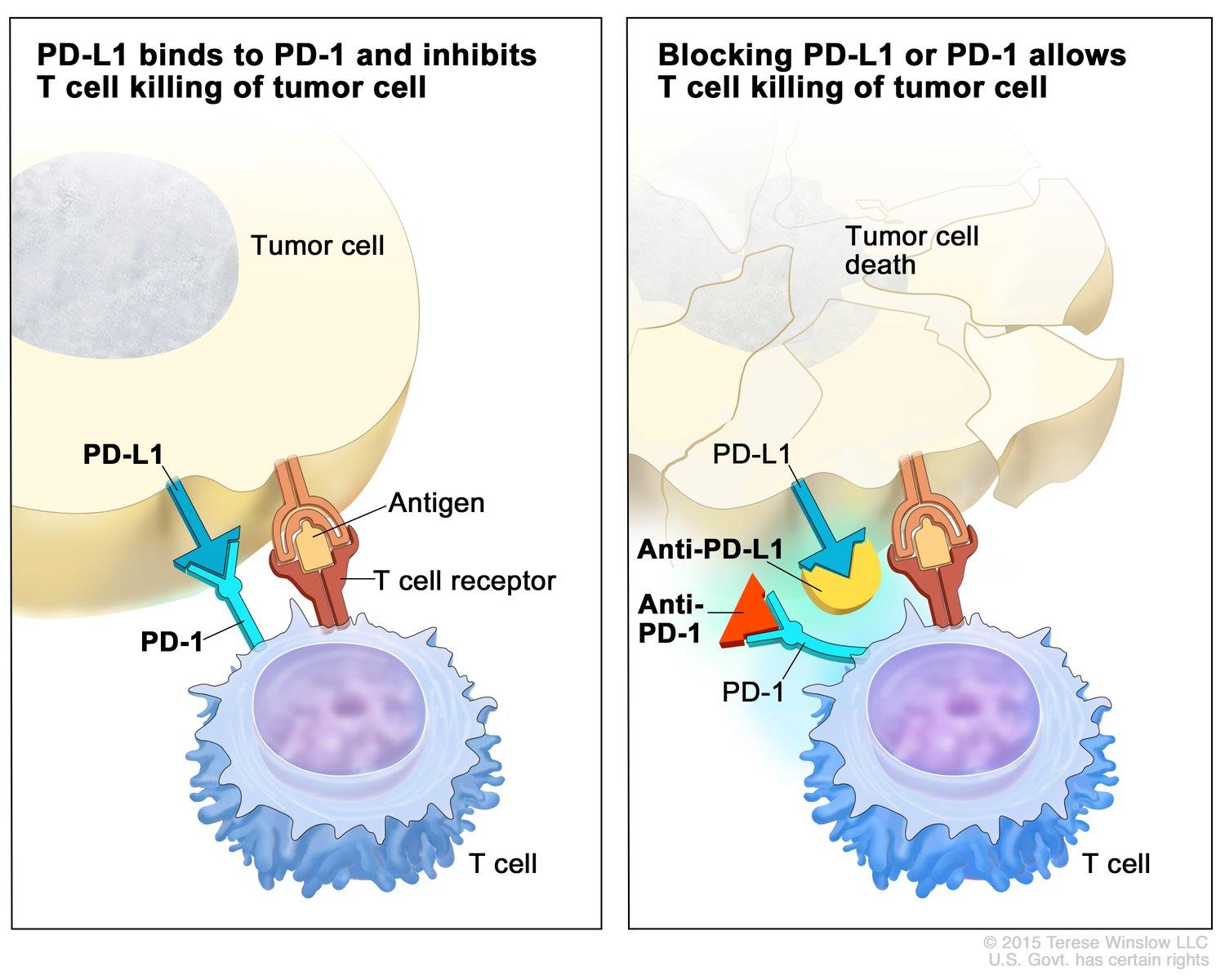}
    \caption{Importance of spatial analysis in evaluating the effectiveness of ICI therapy in killing tumor cells \cite{haanen2015immune}.}
    \label{fig:spatial_syanpse}
\end{figure}

This functionality can be further expanded to generate illustrations representing biological behaviors across various scales in the human body, encompassing interactions between organs and even at the cellular level. For example, Figure \ref{fig:spatial_syanpse} shows two diagrams contrasting the importance of immune ICI therapy in two spatial contexts. The left diagram shows a T cell that cannot destroy the cancer cell due to the spatial association between the tumor and the immune cell. The diagram on the right illustrates a killer T cell that can kill cancer cells due to a blockade (such as anti-PD-L1) at the checkpoint gate (between PD-1 and PD-L1). Existing visualization requires training individuals with a biology background, who may need extensive manual efforts to emphasize the proper biological processes while delivering an easy-to-understand message. Therefore, it is worth investigating whether spatially-aware AI models can represent similar biological behaviors for creating diagrams and examples that are more accessible and interpretable to a broader audience.

\section{Discussion: Bridging the Physical and Digital Space with Atlas-EHR}\label{applications_EHR}
The successful results of research on Atlas-EHR hold enormous potential for various applications. Advanced spatial computing techniques might be used in precision medicine to fight complex diseases (e.g., cancer) by combining genomic data (at a fine biomedical scale) with other EHR data of patients to create customized treatment plans (e.g., targeting particular genetic mutations) and enhance informed decision-making to maximize clinical outcomes for each individual. Spatial computing methods could support real-time monitoring of the body’s inner space to manage chronic disease (e.g., diabetes and heart disease) and predict abnormal events (e.g., high blood pressure, blood clots), enabling doctors to respond quickly to changes in patient’s conditions. Spatiotemporally organized 4D map-based-simulation could enhance the existing EHR platform for grouping and managing patients’ medical history (e.g., lab results, medications, radiology images) based on spatial information in inner spaces such as the human body organs to facilitate more efficient and effective querying and visualizing of relevant information to improve prognostics and treatment planning. Spatially-aware surgery could provide surgeons more in-depth understanding of patient-specific anatomy and physiology, in which inner space, together with a simulation system, would allow them to plan and visualize complex surgeries with greater precision, reducing the need for invasive procedures and minimizing the risk of surgical complications. These are but a few examples of our vision. In short, Atlas-EHR has the potential to revolutionize the healthcare industry. In the following, we aim to highlight some of the specific examples with a diverse range of medical fields. 

\textbf{Examples Across Diverse Medical Fields.} In Orthopedics, Atlas-EHR can provide a detailed 3D models of a patient's anatomy. This could include an in-depth analysis of fractures, deformities, or joint issues from all angles prior to surgery. Surgeons can use these models to precisely plan the location of incisions, implant sizing, bone resections, etc. Moreover, these maps could be used for patient education by visualizing injuries and describing proposed treatments to help patients better understand their condition. Seeing personalized anatomical models improves patient engagement in shared decision making. Atlas-EHR can leverages motion sensors to track patient movement and progress through post-op rehab exercises. This helps clinicians ensure patients are recovering properly after orthopedic procedures. 

Traditional ear-nose and throat (ENT) assessments rely on limited 2D imaging techniques like X-rays and CT scans. While these provide valuable snapshots, they lack the depth perception and real-time interaction crucial for understanding complex ENT structures and their functions. The conversion of 2D scans into interactive 3D models enable doctors to virtually explore intricate ENT structures like the sinuses, inner ear, and vocal cords. This provides a deeper understanding of spatial relationships and potential abnormalities compared to static images. For intricate ENT procedures involving the sinuses, larynx, or inner ear, Atlas-EHR holds immense potential. Imagine surgeons virtually exploring a patient's unique sinus anatomy before surgery, planning implant placement for complex facial reconstruction, or visualizing vocal cord movements in real-time to diagnose voice disorders. 

Another application of Atlas-EHR includes cardiology. Individual variations in heart sizes, shapes, and vessel pathways are crucial for successful procedures. Atlas-EHR can create patient-specific 3D models reconstructed from their unique medical data, providing an intimate virtual replica of their cardiovascular system. Atlas-EHR allows surgeons to plan and rehearse coronary stent insertions within the patient's virtual heart model. This simulated environment enables them to optimize stent size, placement, and approach, minimizing risks and potential complications during the actual procedure.

\section{Conclusion \& Future Work} \label{conclusion_FW}
In the light of the success of applications like Google Maps and Google Earth to communicate big data (e.g., satellite imagery, census) to a broad audience expeditiously, we argue that spatial computing has a similar role to play in the biomedical domain. To this end, we envision Atlas-EHR, a spatially-organized EHRs data for reducing the time needed by healthcare to understand patient history via the next generation of biomedical decision support. We discuss a few challenges and opportunities for five broad areas of spatial computing research to explore the human body. For each domain, we listed some of the recent accomplishments and then highlighted research needs to better solve challenging problems in this space.

For the field of spatial computing, which has been pushing the boundaries of what is possible, the next frontier lies within the human body itself. Exploring inner space computationally will require new thinking and new approaches. ERHs contain a wealth of data that is invaluable to biomedical research, but we need to rethink how we store and extract this data to its full potential. Atlas-EHR provides a conceptual roadmap for the future of spatial computing within the healthcare industry, enabling us to store and analyze EHR data using spatial computing principles. By doing so, we can unlock new insights into complex diseases and develop tailored treatment plans that will maximize patient outcomes.

In our future work, we plan to provide a detailed analysis of challenges related to data collection, retrieval, management processes, and ethical considerations. Challenges that is associated with a complete data life cycle for collecting comprehensive and detailed user data to construct an accurate personal profile. The complexities surrounding the formulation of data retrieval queries, such as how changes in the human body are propagated into the system and affect information retrieval. We also plan to investigate on ethical concerns and the potential for misuse of the system. This may include the possibility of inherent biases within the system, potentially favoring certain demographics over others (e.g., disparities based on body type, gender, or unique physical characteristics), without suggesting corrective measures. The risk of data breaches leading to the denial of insurance claims. Finally, we will extend to an examination of the capabilities of Atlas EHR in facilitating the integration of intermediary imaging modalities, including light sheet microscopy \cite{olarte2018light}, micro-CT \cite{ritman2011current}, and electron microscopy (EM) \cite{koster2003electron}, thereby enhancing the comprehensiveness and precision of diagnostic processes.

\section*{Acknowledgments.}
This material is based upon work supported by the NSF under Grants No. 1901099, and 1916518. 
We also thank Kim Koffolt and the Spatial Computing Research Group for their valuable comments and refinements.

\bibliographystyle{ACM-Reference-Format}
\bibliography{bib}

\appendix

\section{Overview of Existing EHR Platform}
\begin{figure}[ht]
    \centering
    \includegraphics[width=\linewidth, height=10cm]{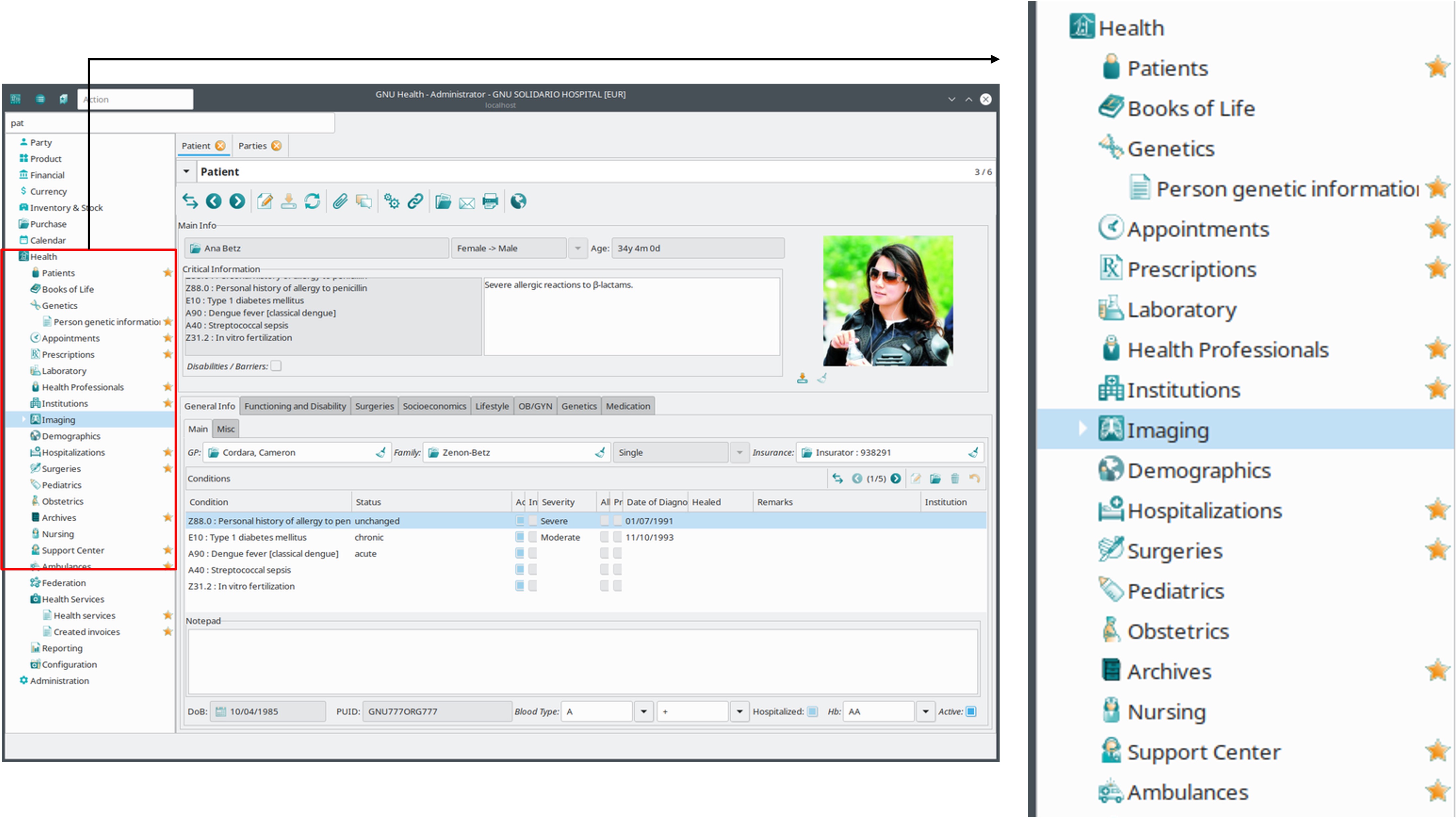}
    \caption{An existing EHR software without a spatially-defined taxonomy}
    \label{old_ehr}
\end{figure}

Major electronic health record (EHR) systems, including Epic Systems, contain extensive patient histories that span years or even decades of medical data. These detailed records present healthcare professionals with the challenge of efficiently navigating massive amounts of complex information (see Figure \ref{old_ehr} for an example EHR interface). The current temporal format of EHRs makes it difficult for medical staff to uncover and analyze the intricate connections between diverse types of patient data gathered by various sensing technologies across multiple scales and time periods. For instance, physicians may struggle to identify correlations between whole-body scan results from years ago and cellular-level test results from a recent visit. The extensive breadth and depth of EHRs, which contain measurements ranging from macro-level bodily systems down to micro-level cellular components, compounds this difficulty.

\section{Spatial Computing in Geographic Space}\label{GIS}
This section provides an overview of some state-of-the-art techniques in five broad areas of spatial computing as they have been applied to geographic space. At the end of each discussion, we note a corresponding challenge that will need to be addressed at the scale of inner space. A full discussion of these challenges and the research needed to meet them is in Section \ref{inner_space}.


\subsection{Spatial Database Management} 
A database management system (DBMS) is a computerized platform that serves as an interface between a database and its end users, allowing users to retrieve, update, and manage how the information is organized and optimized. The traditional database provides persistence, concurrency control, and scalability. Persistency is achieved by storing data on disk, allowing it to survive system failures, and ensuring data integrity and consistency. Concurrency control is achieved by ensuring that multiple transactions simultaneously executing do not interfere with each other and maintain data consistency. Lastly, scalability is the ability to handle big data without a decrease in performance through techniques such as indexing, partitioning, and sharding. While a traditional DBMS is efficient for non-spatial queries, such as student summary statistics, it is not efficient for spatial queries, like listing the name of cells in a spatial neighborhood (e.g., 50 pixels distance) of a tumor cell. In order to facilitate the use of spatial data (e.g., spatial transcriptomics data and cellular maps), we need spatial DBMS.

A spatial database management system is specifically designed to store, manage, and analyze geographical or spatial data. This data includes geographic coordinates, shapes, and locations, and is used to represent real-world objects such as roads, buildings, and geographical features. A spatial DBMS provides spatial data models (e.g., OGC's simple features), spatial abstract data types (ADTs), and a query language from which these ADTs are callable. It supports spatial indexing (e.g., space-filling curves such as Z-curve), and provides efficient algorithms (e.g., R, R+, R* trees) for spatial operations and ad-hoc queries. 

The input to these spatial indexing algorithms is mainly spatial objects (e.g., city boundaries and counties) that are rigid and pre-defined in geographic space. However, spatial objects in inner space, including human organs if affected by diseases, are non-rigid and change in shape, size, texture, and function. Thus, an important question is how one defines the input MOBRs to spatial indexing techniques (e.g., R, R+, R* trees). We elaborated more in Section \ref{inner_space}. 

\subsection{Spatial Pattern Mining} 
\begin{figure}[ht]
    \centering
    \includegraphics[width=0.8\textwidth]{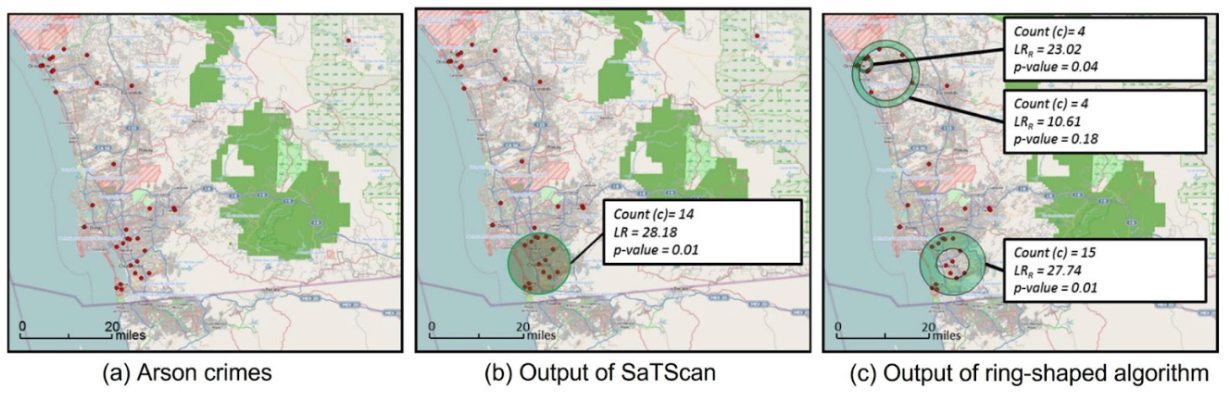}
    \caption{(a)  33 arson crimes in San Diego in 2013. Red dots represent locations of crimes. (b) Comparison of SaTScan and  (c) ring-shaped hotspot detection on the same arson crime data. Green circles/rings show the output of both algorithms.}
    \label{fig:ring-shaped}
\end{figure}

Spatial pattern mining aims to discover potentially useful, interesting, and non-trivial patterns from spatial datasets. One of the most common pattern families is geospatial hotspot detection. The goal is to identify often pre-defined regions (e.g., circular, rectangular) that are denser than surrounding areas. However, in some domains, activities may occur in a ring-shaped pattern. A recent effort to model such behaviors is ring-shaped hotspot detection \cite{eftelioglu2014ring}, in which the intuition is serial criminals who often commit crimes neither too close to their home nor too far away. For example, Figure \ref{fig:ring-shaped} (a) shows 33 arson crimes in San Diego in 2013. Ring-shaped hotspot detection outputs several ring patterns (Figure \ref{fig:ring-shaped} (c)), whereas SaTScan outputs one large circular area (Figure \ref{fig:ring-shaped} (b)). More recent approaches have also explored geospatial hotspot detection over a spatial network (e.g., linear hotspot detection \cite{LinearTang2017}) as well as statistically arbitrary shape hotspots (e.g., sig-DBSCAN \cite{XieSDBSCAN19}). 


Another common pattern family is spatial co-location pattern detection. Spatial co-location patterns represent spatial features whose instances are
located near one another given a neighbor relation $R$. The most commonly used interest measure in co-location analysis is the participation index. A participation index is defined as the minimum participation ratio (PR) of the features in a co-location pattern. The participation ratio of a feature $f_j$ in a co-location pattern $C$ under input distance threshold $d$, ${\text{PR}(C,f_j,d)}$, is the fraction of objects of the feature participating in instances of the pattern. While  many studies focus on global co-location patterns, some patterns may only exist at a regional level, and more recent approaches \cite{LocalLi2018, DBLP:conf/gis/GhoshGSAS22} aim to explore such behaviors. Table \ref{tab:co-location_usecases} lists a few regional co-location application domains and their example use cases. 


A spatially-explainable classifier representing the interactions between immune and tumor cells can be created using a hand-crafted measure derived from co-location analysis, specifically the participation index. However, this measure is used in isotropic space, with the same intensity regardless of measurement direction, which may not be enough to capture relevant spatial interactions, such as \textit{surrounded by}, which could be biologically significant. Thus, an important question is how one defines new spatial concepts beyond open geospatial consortium's (OGC) simple features to describe co-location patterns of cells. 
 We elaborate on this issue and  similar challenges in Section \ref{inner_space}.


\begin{table}\scriptsize
\footnotesize
\centering
\caption{Regional colocation applications}
\label{ApplicationDomain}
\begin{tabular}{p{2.0cm}p{10cm}p{5cm}}
\hline
Application Domain  &  Example Use Cases\\ 
\hline
Oncology   & <Reponsder, Cytotoxic T-cell and Cancer cell>, <Non-responder, Regulatory T-cell, Tumor cells>\\ 
\hline
Retail  & <China, McDonald’s and KFC>, <USA, McDonald’s and Jimmy John’s>\\ 
\hline
Public Safety & <Region around bars, Assault crimes, and drunk driving>\\ 
\hline
Transportation  & <Near bus depots, High $NO_{x}$ concentrations, and buses>\\
\hline
\end{tabular}
\label{tab:co-location_usecases}
\end{table}


\subsection{Positioning, Navigation, Location-based Services}Positioning is the process of determining an object's orientation and location in space using global navigation satellite systems (GNSS), such as the global positioning system (GPS), which accurately calculates an object's location, velocity, and time. In this context, triangulation \cite{hartley1997triangulation} by measuring the angles between three points and trilateration \cite{thomas2005revisiting} by measuring the distances to three or more points are well-established techniques to determine the unknown location of an object. The widespread availability of GPS systems is primarily due to their low-cost and compact design, which is facilitated by their use of very-large-scale integrated (VLSI) circuit implementations. This has further led to GPS systems being easily incorporated into mobile phones and tablets, followed by ubiquitous location-based services and numerous applications, including routing and navigation, ride-sharing, surveillance, and context advertising. 

Even with worldwide availability, GPS signals are largely unavailable indoors due to being blocked by buildings and other structures, where we humans spend  80\% to 90\% of our time \cite{greed2004inside}. Hence, indoor navigation within enclosed structures such as buildings, alleys, or other closed underground locations (e.g., basements) has become an emerging research field. The WiFi positioning system (WPS) uses hotspots and other access points to determine device location. It is often used in conjunction with GPS for better accuracy. Time geography \cite{miller2005measurement} is also used with WiFi traces to quantify space-time uncertainty where WiFi hotspots are in low capacity resulting in very low location precision. Other positioning systems such as Bluetooth Low Energy (BLE) beacons, and ultrasound or infrared sensors are also emerging and current research trends. 


Although various studies \cite{grood1983joint, morosan2001human} have attempted to develop a coordinate system for specific areas of the human body, such as the brain, relatively little effort \cite{rood2019toward} has been devoted to creating a comprehensive spatial reference system for the entire body akin to the latitude/longitude system used in geographic space. These studies have employed a range of techniques, including anatomical landmarks, motion capture, and imaging (e.g., MRI data), to establish coordinate systems for localized regions of the body. Nevertheless, the fundamental question remains: how can a stable and standardized spatial reference system be defined given the intricate and dynamic nature of the human body, with its thousands of anatomical structures that can shift and change over time? We will elaborate on this more in Section \ref{inner_space}.

\subsection{Sensing} An important question in geography is what has been the impact of climate change, urbanization, and population growth on forest cover in recent decades? Geographers traditionally answered this question through labor-intensive manual surveys, which were limited to small geographic areas. Now, remote sensing satellites (e.g., Landsat) allow global monitoring and exploration of land cover and spatial patterns such as drought and weather forecasts. With unmanned aerial vehicles (UAVs), multispectral and hyperspectral sensors, remote sensing data is collected at even higher resolutions to study plant diseases, physiology, and crop growth under extreme weather conditions.


In recent years, many societally important applications in the remote sensing field, including but not limited to detecting land use and land cover, controlling forest fires, wetland management, and observing climate changes, have gained full attention with the rise of deep neural networks (DNNs) for tasks ranging from image classification to semantic segmentation \cite{cecotti2020grape, yuan2020deep}. DNN techniques have addressed the limitations of timely and expensive hand-crafted feature construction while providing better performance on selected metrics (e.g., accuracy). However, post-classification processing via rule-based software (e.g., e-cognition) may be often required to provide more interpretable results and reduce noise due to the probabilistic nature of these techniques. 

In addition to deep neural network (DNN) techniques, several other data-driven approaches have been developed to address issues such as salt and noise errors due to the self-correlated nature of spatial data. One recent technique is the spatial decision tree \cite{DBLP:conf/icdm/JiangSZKC13}, which reduces classification error by considering both local and focal (i.e., neighborhood) information to determine the traversal direction of a sample point. These models are typically designed to handle well-structured data, such as raster data representing phenomena over a continuous space. However, they are inadequate for dealing with simple yet critical geometric structures, such as cellular maps derived from MxIF imagery. DNN-based techniques (e.g., PointNet++ \cite{qi2017pointnet++}, DGCNN \cite{wang2019dynamic}) have been proposed to address these limitations. However, these models are not specifically designed to handle multi-categorical point sets (e.g., cellular maps with different types), which do not take full advantage of the spatial relationships between different categories of points.

\subsection{Cartography and Visualization} 
\begin{figure}[ht]
    \centering
    \includegraphics[width=0.7\textwidth]{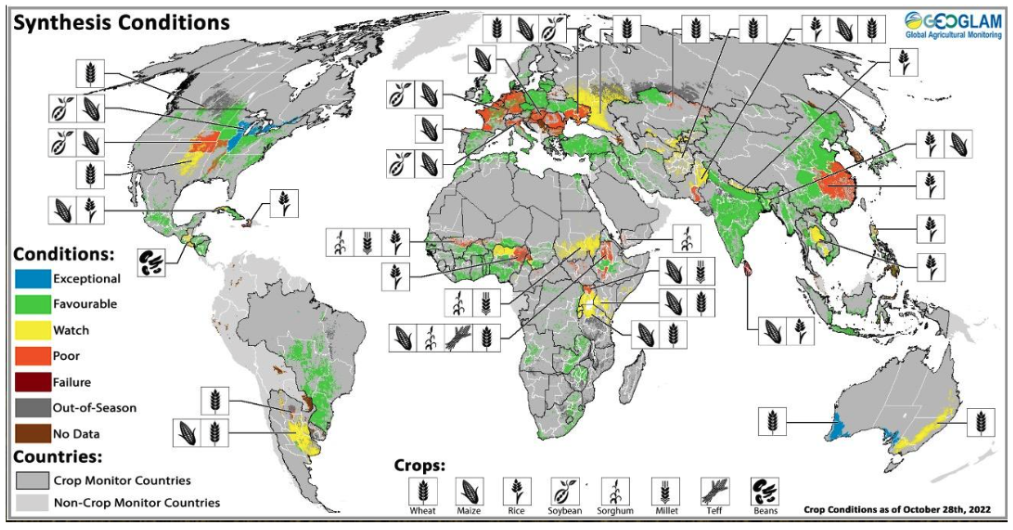}
    \caption{Periodic Global Agricultural Monitoring from GEOGLAM as of October 2022}
    \label{fig:geoglam}
\end{figure}

A geographic information system (GIS) is an integrated platform to collect, manage, analyze, and visualize spatiotemporal information. Many such systems, including ArcGIS and QGIS, are widely used in application areas such as public health, public safety, and agriculture. These applications provide a user-friendly interface to import data from third parties (e.g., the US Census), use spatial statistics techniques (e.g., inverse distance weighted, Kriging, etc.) for interpolation, and visualize derived patterns (e.g., air quality interpolation) in maps. In this context, GEOGLAM \cite{DBLP:conf/igarss/Becker-ReshefJW18}, the Group on Earth Observations Global Agricultural Monitoring, with an inherently synthetic focus, is a collaborative initiative to leverage and build upon existing programs and activities through coordinated Earth observation, capacity development, monitoring, and research and development activities. A critical research outcome from GEOGALM is crop monitoring for AMIS (wheat, maize, rice, and soybean) for early warning systems based on multi-source, consensus assessments of the crop growing conditions, status, and agro-climatic conditions (See Figure \ref{fig:geoglam}).

To further improve the functionality and capacities of GIS software, much research has been conducted on integrating query processing, indexing, and optimization techniques to deal with ever-increasing volumes of spatial big data along with enhancing classification/detection performance by incorporating deep learning techniques (e.g., CNNs). One example is the integration of spatial modeling techniques, such as network analysis, into GIS software. This type of integration allows for the analysis and simulation of spatial relationships, such as the movement of goods and people along transportation networks, in order to make informed decisions about the design and operation of these networks. 

As with GIS software, the recent explosion in spatial pathology data has led to the development of many multiplexed commercial software tools (e.g., IonPath, NanoString) that incorporate machine learning and image processing techniques for identifying multiple biomarkers and their corresponding locations. Interested readers can refer to a comprehensive survey of these technologies for further details \cite{tan2020overview}. However, while existing morphologic or count-based approaches have shown some success, limited efforts have been made towards integrating advanced spatial data science techniques, such as spatially-explainable models and spatially generative artificial intelligence, to visualize co-localized biomarkers and discovering complex spatial relationships beyond co-location. We elaborate on this issue further in Section \ref{inner_space}.

\end{document}